\pdfoutput=1
\documentclass[11pt]{article}

\usepackage[]{EMNLP2022}
\usepackage{times}
\usepackage{latexsym}

\usepackage[T1]{fontenc}
\usepackage[utf8]{inputenc}

\usepackage{microtype}

\usepackage{inconsolata}

\usepackage{times}
\usepackage{soul}
\usepackage{url}
\usepackage{graphicx}
\usepackage{amsmath}
\usepackage{amsthm}
\usepackage{booktabs}
\usepackage{algorithm}
\usepackage{algorithmic}

\usepackage{xspace} 
\usepackage{diagbox}
\usepackage{amssymb} 
\usepackage{multirow}
\usepackage{xcolor}
\usepackage{xspace} 
\usepackage{color} 
\usepackage{bm}
\usepackage{colortbl}
\usepackage{setspace} 
\usepackage{array}
\usepackage{arydshln} 
\usepackage[normalem]{ulem}

\newcommand{\apicoder}{APICoder\xspace}
\newcommand{\apiretriever}{APIRetriever\xspace}
\newcommand{\pandaseval}{PandasEval\xspace}
\newcommand{\numpyeval}{NumpyEval\xspace}
\newcommand{\monkeyeval}{MonkeyEval\xspace}
\newcommand{\beatnumeval}{BeatNumEval\xspace}
\newcommand{\torchdataeval}{TorchDataEval\xspace}
\newcommand{\codegen}{\textsc{CodeGen}\xspace}
\newcommand{\codegenmono}{\textsc{CodeGen-Mono}\xspace}
\newcommand{\codegenapi}{\textsc{CodeGenAPI}\xspace}
\newcommand{\yes}[1]{{\color{ForestGreen}{\cmark}}}
\newcommand{\no}[1]{{\color{red}{\xmark}}}

\newcommand{\zan}[1]{\textbf{\textcolor{blue}{Zan: first refine to here}}}

\title{When Language Model Meets Private Library}

\author{Daoguang Zan$^{1,2}$\thanks{~~Work done as an intern at Microsoft Research Asia.}, Bei Chen$^3$, Zeqi Lin$^3$, Bei Guan$^{2,4}$, Yongji Wang$^{2,4,5}$, Jian-Guang Lou$^3$ \\
  $^1$Cooperative Innovation Center, Institute of Software, Chinese Academy of Sciences \\
  $^2$University of Chinese Academy of Sciences;
  $^3$Microsoft Research Asia \\
  $^4$Integrative Innovation Center, Institute of Software, Chinese Academy of Sciences \\
  $^5$State Key Laboratory of Computer Science, Institute of Software, Chinese Academy of Sciences \\
  \texttt{\{daoguang@, guanbei@, ywang@itechs.\}iscas.ac.cn} \\
  \texttt{\{beichen, zeqi.lin, jlou\}@microsoft.com} \\
}

\begin{document}
\maketitle
\begin{abstract}
With the rapid development of pre-training techniques, a number of language models have been pre-trained on large-scale code corpora and perform well in code generation. In this paper, we investigate how to equip pre-trained language models with the ability of code generation for private libraries. In practice, it is common for programmers to write code using private libraries. However, this is a challenge for language models since they have never seen private APIs during training. Motivated by the fact that private libraries usually come with elaborate API documentation, we propose a novel framework with two modules: the \apiretriever finds useful APIs, and then the \apicoder generates code using these APIs. For \apiretriever, we present a dense retrieval system and also design a friendly interaction to involve uses. For \apicoder, we can directly use off-the-shelf language models, or continually pre-train the base model on a code corpus containing API information. Both modules are trained with data from public libraries and can be generalized to private ones. Furthermore, we craft three benchmarks for private libraries, named \torchdataeval, \monkeyeval, and \beatnumeval. Experimental results demonstrate the impressive performance of our framework.\footnote{Our work is available at \url{https://github.com/microsoft/PyCodeGPT/tree/main/apicoder}.}. 
\end{abstract}

\section{Introduction}\label{introduction}
Code generation, automatically generating code snippets based on user descriptions, is one of the long-standing challenges in the software engineering and artificial intelligence communities. With the rapid development of pre-training techniques, a number of language models are pre-trained on large-scale code corpora and able to generate decent code snippets, for example, Codex~\cite{codex}, AlphaCode~\cite{alphacode}, \codegen~\cite{codegen}, and InCoder~\cite{incoder}. They bring fresh energy to code generation and improve coding efficiency~\cite{evaluation_codex,codet}. Although making remarkable progress, these models may be biased towards generating code that is similar to the training distribution~\cite{codex}. What if one wants to generate code beyond the training distribution? A real-world scenario for programmers is to write code using a private library, which is very common in practice. For example, for security and functionality reasons, companies often build private libraries for internal use only. Private libraries provide a number of private APIs that have not been seen by the language models and are also not publicly available on any code hosting platform like GitHub. Therefore, it is worth exploring whether and how pre-trained language models can generate code using private libraries. 

\begin{figure}[t]
    \small
    \centering
    \includegraphics[width=1.0\linewidth]{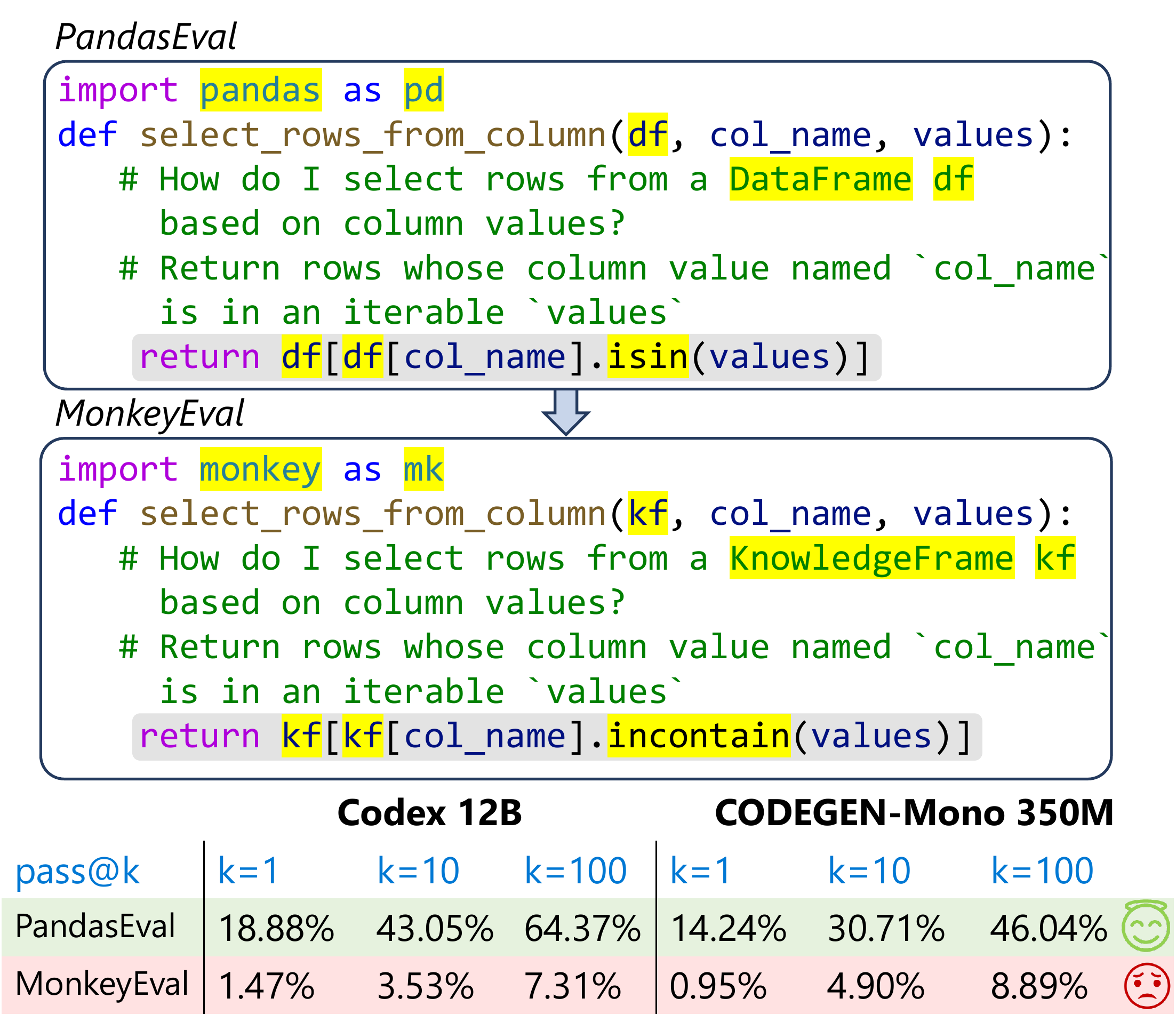}
    \caption{An example of converting PandasEval (public) to MonkeyEval (private). The changed parts are highlighted in yellow. The performance of Codex $12$B and \codegenmono $350$M is shown at the bottom.}
    \label{fig:figure1}
\end{figure}

It is challenging for existing language models to generate code that uses a private library directly. A practical evidence is shown in Figure~\ref{fig:figure1}. We built a pseudo private library named Monkey based on a public one named Pandas. \pandaseval~\cite{cert} is a benchmark consisting of $101$ Pandas programming problems. We convert all Pandas-related keywords in \pandaseval into the new version and construct \monkeyeval (details in Section \ref{sec:benchmark}). As seen in Figure~\ref{fig:figure1}, Codex $12$B and \codegenmono $350$M show a significant drop in performance on the private \monkeyeval compared to their performance on the public \pandaseval. For example, Codex $12$B drops from $18.88$\% to $1.47$\% on pass@$1$, showing the inadequacy of the language models in code generation for private libraries.

To meet the challenge, we propose a framework to equip pre-trained language models with the ability to generate code that uses private libraries. As is known, private libraries usually come with elaborate API documentation, which motivates our main idea to mimic the process of a programmer learning to write code using private libraries. This process is also known as API practices in the software engineering field~\cite{apipractices}: first learning the private API documentation and then invoking the APIs to implement the needed functionalities. Analogically, there are two modules in our framework: an \apiretriever first retrieves the useful APIs based on the programming problem and the API documentation, and then an \apicoder uses these APIs to generate code. For \apiretriever, we train a dense retriever and also design a friendly interaction to involve users in the loop optionally. For \apicoder, we can directly use existing language models of code generation, such as \codegen, to invoke the private APIs; furthermore, to better teach a language model how to invoke APIs, we also continually pre-train the base model on a code corpus containing API information from public libraries, and obtain our reinforced model called \codegenapi. Since we only have access to the data of public libraries during training, we expect that \apiretriever and \apicoder can be generalized to private libraries via learning.

To evaluate the code generation for private libraries, we craft three benchmarks, named \torchdataeval, \monkeyeval, and \beatnumeval. \torchdataeval includes $50$ programming problems using the TorchData library. The last two are adapted from \pandaseval and \numpyeval~\cite{cert}, respectively, each consisting of $101$ programming problems. Extensive experiments on the three benchmarks have revealed that our framework effectively improves the performance of pre-trained language models on code generation for private libraries. We also provide a thorough analysis to facilitate progress in this direction.

\begin{figure}[t]
    \small
    \centering
    \includegraphics[width=1.0\linewidth]{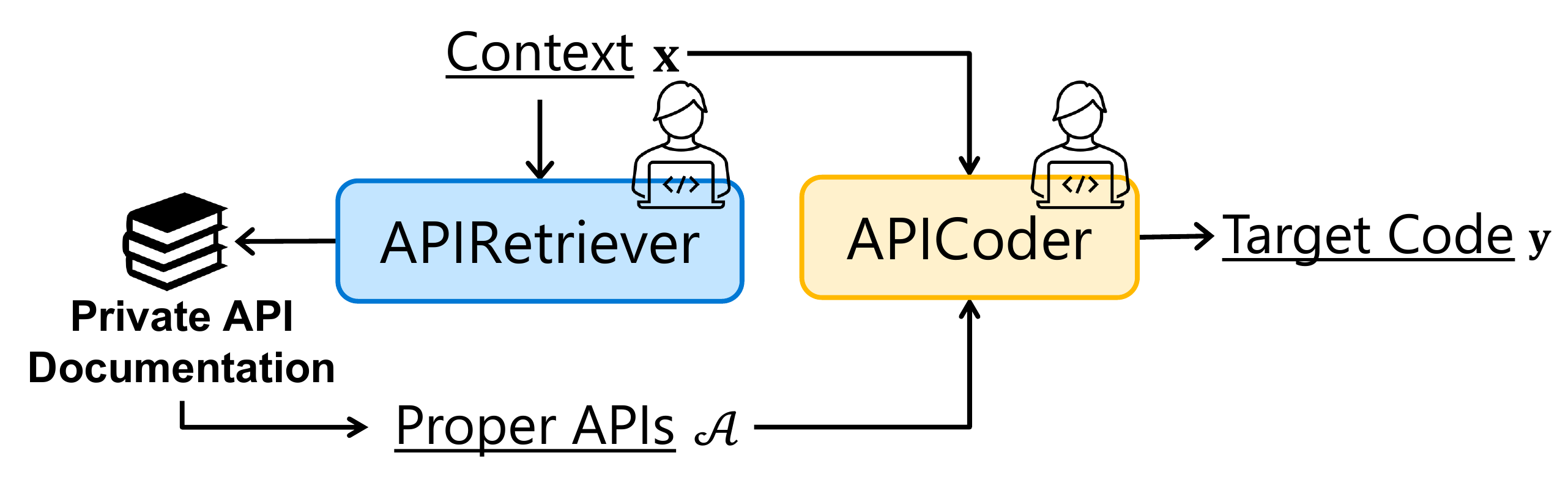}
    \caption{The overview of our proposed framework.}
    \label{fig:apihelper}
\end{figure}

\begin{figure*}[t]
    \small
    \centering
    \includegraphics[width=1.0\linewidth]{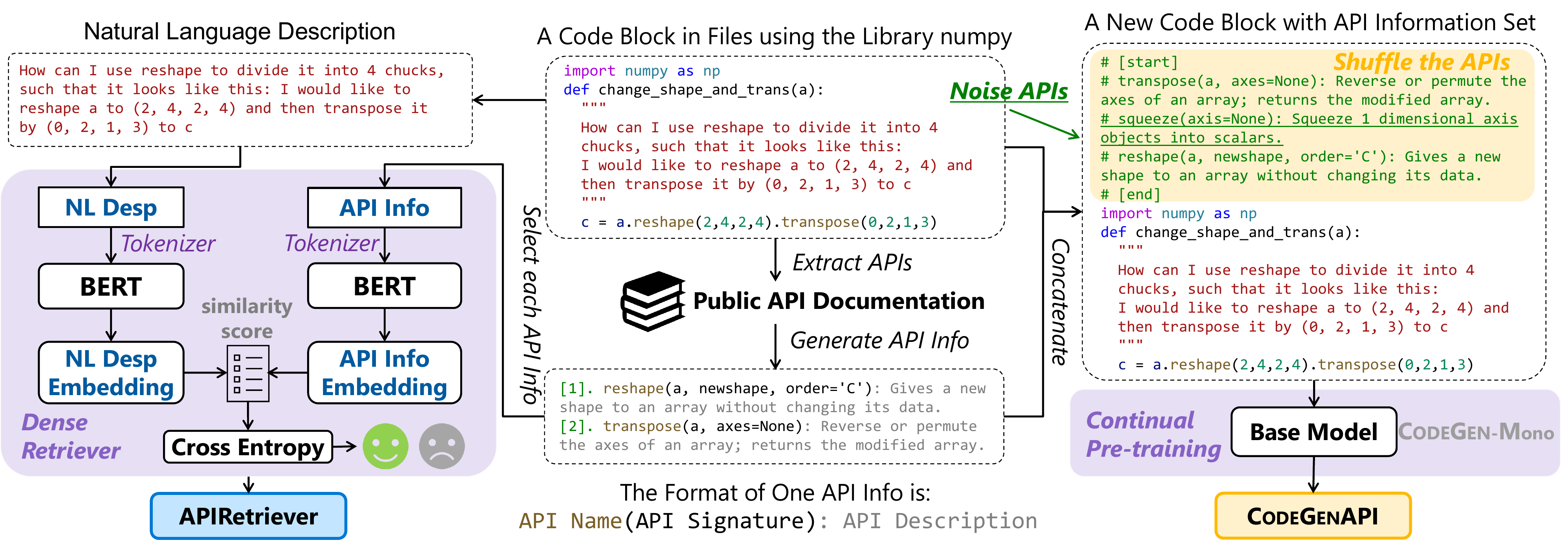}
    \caption{The training process of \apiretriever and \codegenapi.}
    \label{fig:apiretriever_apicoder}
\end{figure*}

\section{Framework}
First, we would like to define the task of code generation formally. Given \textit{context}, the task aims to generate \textit{target code}. In Figure~\ref{fig:figure1}, context and target code are shown in white and grey backgrounds, respectively. Context consists of a comment, which is a natural language description of the programming problem, and a code snippet including import statements, function header, etc. Target code solves the programming problem in context. We denote the context by $\mathbf{x}$. Code generation model $\mathcal{M}$ outputs target code $\mathbf{y}$ based on $\mathbf{x}$. For the task of code generation for private library, the context $\mathbf{x}$ contains the instruction for using a private library, such as an import statement. The target code $\mathbf{y}$ contains the calls of the corresponding private library APIs.

As mentioned in Section~\ref{introduction}, private libraries are usually equipped with elaborate API documentation. As a technical reference manual outlining how to use the library, API documentation typically includes a quick start guide, tutorials, and an instruction for each API (e.g., API name, signature, description, parameters, and examples). To take advantage of the API documentation, we propose to mimic the generic process of a programmer coding with private APIs, and design a framework to generate code that can invoke private APIs. The framework consists of \apiretriever and \apicoder with the overview shown in Figure~\ref{fig:apihelper}. Given the context, \apiretriever $\mathcal{M}_{\rm{R}}$ aims to retrieve possible used APIs from the API documentation; and \apicoder $\mathcal{M}_{\rm{C}}$ is dedicated to generating code using the retrieved APIs. The process can be formalized as $\mathcal{A}=\mathcal{M}_{\rm{R}}(\mathbf{x})$ and $\mathbf{y}=\mathcal{M}_{\rm{C}}(\mathcal{A};\mathbf{x})$, where $\mathcal{A}$ represents the set of information of all proper APIs, and each $\mathbf{a}\in\mathcal{A}$ is the information of an API. In our implementation, we design the API information to include the API name, signature and description. Note that we only use the first sentence of the API description since it is sufficient to summarize. 

\section{Methodology}
We have introduced our framework that provides pre-trained models a fantastic way to deal with private libraries. 
In this section, we present the data collection, followed by the detailed design of our \apiretriever and \apicoder.

\subsection{Data Collection} \label{sec:data}
We collect API information and code files of public libraries due to the fact that we can only access data from public libraries. Then we train the models based on the public data with the expectation that the model can be generalized to private libraries. For API information, we consider the $31$ most popular public libraries in Python (e.g., Pandas, NumPy, and scikit-learn) according to the popularity ranking on StackOverFlow\footnote{\url{https://stackoverflow.com/tags?tab=popular}}. For each of the libraries, we crawled its API documentation and extracted detailed information about each API, including the API name, signature, description, parameters, usage examples, and so on. Please refer to Appendix~\ref{sec:31lib} for the details of the 31 public libraries. For code files, we first collected a $330$GB corpus from GitHub containing $60.6$M python files and then extracted those files that involved one or more API calls from the $31$ public libraries. After a bunch of pre-processing strategies, for example, de-duplicating, cleaning, and formatting, we finally obtained $4.54$M python files, denoted by $\mathcal{D}$. 

\subsection{\apiretriever}
\apiretriever aims to find the proper APIs based on the description of a programming problem. We regard it as a dense retrieval task~\cite{rocketqa,xiong2020approximate,santhanam2021colbertv2,s2ql,formal2022distillation} and design a simple dual-encoder model~\cite{dense-retrieval} to retrieve the possible used APIs for each programming problem. To further boost the retrieval performance, a friendly interaction approach is designed to involve users.

\paragraph{Training.}
To train \apiretriever, we need a large amount of pairwise data, natural language description and API information. We first segment each python file $\mathbf{d}\in\mathcal{D}$ into $K$ code blocks $(d_1,d_2,{\cdots},d_K)$ using the pip-tools, i.e., redbaron, autopep8, and docformatter, where each code block is a relatively well-rounded code fragment, such as a function or a class. For each code block $d_i$, we extract all API names and obtain the corresponding API signatures and descriptions by searching our collected $31$ API documentations\footnote{If an API name matches more than one candidate API, we randomly pick one.}. The information of an API includes its name, signature and description, denoted by $\mathbf{a}\in\mathcal{A}$. Each $\mathbf{a}$ and the natural language description $\mathbf{p}$ extracted from the same code block $d_i$ are regarded as a positive training sample. For the negative training sample, we randomly sample an API $\hat{\mathbf{a}}$ that is unrelated to $d_i$ from the same library. 
In total, we obtained $40.3$M ($\mathbf{p}$, $\mathbf{a}$, $\hat{\mathbf{a}}_1$, $\hat{\mathbf{a}}_2$, \dots) sets as training samples. As in Figure~\ref{fig:apiretriever_apicoder}, the left part shows the training process of \apiretriever. Our \apiretriever is a dual-encoder model. The two dense encoder, $E_{\mathbf{p}}(.)$ and $E_{\mathbf{a}}(.)$, map $\mathbf{p}$ and $\mathbf{a}$ to $z$-dimensional vectors, respectively. Then, we use the dot product of their vectors to calculate the similarity score formalized as $E_{\mathbf{p}}(\mathbf{p})^{\top}E_{\mathbf{a}}(\mathbf{a})$, where $E_{\mathbf{p}}(.)$ and $E_{\mathbf{a}}(.)$ are implemented by two independent BERT~\cite{bert} with base-uncased version. We use BERT instead of CodeBERT~\cite{codebert} as most tokens in $\mathbf{p}$ and $\mathbf{a}$ are natural language rather than programming language.

\paragraph{Inference.} 
After the training phase, we can use \apiretriever to retrieve private APIs for each programming problem description. In detail, we apply $E_{\mathbf{a}}$ to all the APIs and index them by FAISS~\cite{faiss} offline. Given a new programming problem description $\mathbf{p}$ at run-time, we only need to produce its embedding $v_{\mathbf{p}}=E_{\mathbf{p}}(\mathbf{p})$ and recall the top-$k$ APIs with the embeddings closest to $v_{\mathbf{p}}$.

\begin{figure}[t]
    \small
    \centering
    \includegraphics[width=1.0\linewidth]{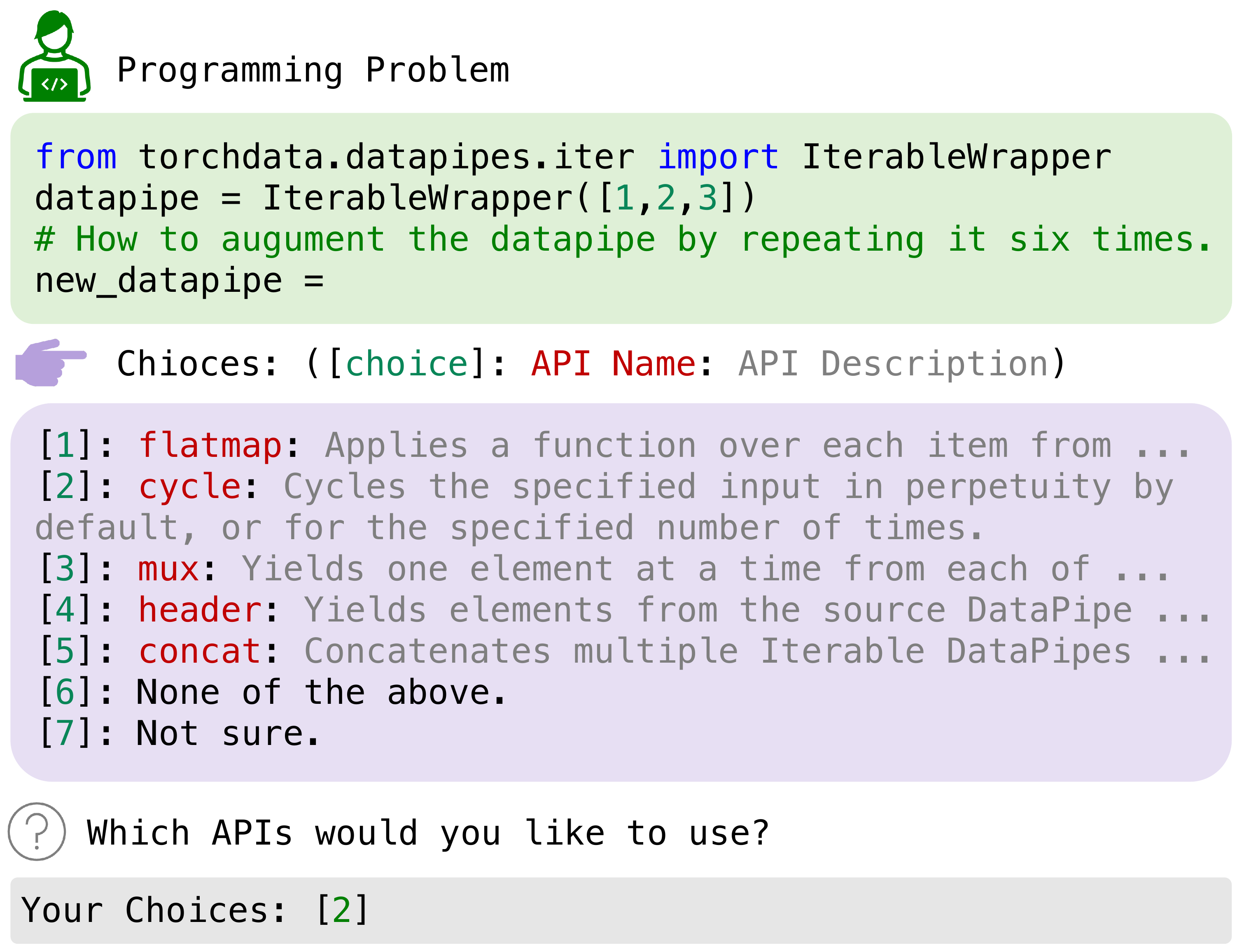}
    \caption{Friendly interaction interface for users.}
    \label{fig:human_in_the_loop_tests}
\end{figure}

\paragraph{Human Interaction with \apiretriever.}
In order to further increase the accuracy of API retrieval, we provide a friendly interaction interface to allow humans in the loop with \apiretriever, as shown in Figure~\ref{fig:human_in_the_loop_tests}. In the interaction interface, we give the programming problem and the top-$5$ APIs retrieved by \apiretriever, and let users choose one or more APIs that may be used in the target code. Note that we only provide API names and descriptions to users, as we find in our empirical experiments that providing API signatures has a negative effect on making the correct choice.

\subsection{\apicoder} \label{sec:apicoder}
\apiretriever finds useful APIs for a programming problem, and then \apicoder aims to generate code that solves the problem with these APIs. We make use of the most straightforward way for \apicoder: prompting API information set $\mathcal{A}$ in front of the context $\mathbf{x}$. Formally, the \apicoder can be written as $\mathbf{y}=\mathcal{M}_{\rm{C}}({\rm{Concat}}(\mathcal{A},\mathbf{x}))$, where ${\rm{Concat}}(\mathcal{A},\mathbf{x})$ means to concatenate the API information set and the context. Examples can be found in Figure \ref{fig:apiretriever_apicoder}. Each API information is in the form of ``\texttt{name(signature):description}''. This is to mimic programmers learning the APIs properly before writing code using them.

Technically speaking, the off-the-shelf code generation models, such as CodeT5, CodeGPT, CodeClippy, CodeParrot, \codegen, and Codex, can be applied directly to land \apicoder. Although these base models can achieve gains in correctly invoking APIs, they have not learned how to use them as an explicit training task. To better use the APIs, we devised a fantastic idea of continually pre-training the base models using code files with API information inserted.

In practice, we use \codegenmono $350$M \cite{codegen} as our base model, based on which we continually pre-train and obtain our reinforced model called \codegenapi. \codegen is a GPT-based model skilled at generating code. We choose it because it is by far the most popular and publicly available model. As for the training corpus, we use the collected python files $\mathcal{D}$ mentioned in Section \ref{sec:data}. Firstly, as done for \apiretriever, each file $\mathbf{d}\in\mathcal{D}$ is split into $K$ code blocks $(d_1,d_2,{\cdots},d_K)$. For each code block $d_i$, we obtain the set of API information $\mathcal{A}_i$. Then, the $K$ code blocks and sets of API information are cross-merged to output a new file $\hat{\mathbf{d}}=(\mathcal{A}_1,d_1,\mathcal{A}_2,d_2,{\cdots},\mathcal{A}_K,d_K)$. This mimics the process of API information as a prompt for each block. Then, we continually pre-train the base model on the new code files, teaching the model to write code based on the prompted APIs. In addition, as shown in Figure~\ref{fig:apiretriever_apicoder}, to make \apicoder more robust, we shuffle the APIs in each set $\mathcal{A}_i$ and also add noise APIs, since \apiretriever does not know the order of APIs in the target code and often finds incorrect APIs.

During the training phase of \codegenapi, unlike the previous settings that force all files to have the same priority, we design a resampling strategy to enable high-quality python files to appear more frequently and vice versa. The strategy considers the star number of the repository, the unit test function rate of the code file, and the API rate of the code file. More details can be found in Appendix~\ref{appendix:resampling_strategy}.

\section{Benchmark Construction} \label{sec:benchmark}
Private libraries are commonly used in practice, but few attempts have been made to evaluate the performance of generating code invoking private APIs. To fill this gap, we craft three benchmarks, called \torchdataeval, \monkeyeval, and \beatnumeval. Each programming problem consists of context, target code, and the corresponding test cases.

To create a realistic benchmark for evaluating code generation for private library, we use TorchData, a Python library released just recently\footnote{Our base model, \codegen, is pre-trained with GitHub data before $2021$-$10$. TorchData was released after this time point and no code files using it are available on GitHub so far; hence we can consider it as a private library.}. We carefully learnt the official API documentation of TorchData and made sure we were proficient in all APIs. Then, we manually created $50$ programming problems based on the API usage examples in the documentation. Two volunteers with extensive experience in Python were invited to check the correctness of each problem. We control the difficulty of the programming problems by the number of APIs in the target code. The percentage of programming problems containing $1$ API, $2$ APIs, and more APIs is set to $6$:$3$:$1$.

We also construct two benchmarks using pseudo private libraries, named \monkeyeval and \beatnumeval, each containing $101$ programming problems. They are modified from \pandaseval and \numpyeval, which were proposed for the public libraries Pandas and Numpy~\cite{cert}. In detail, we manually modified all library-related keywords in \pandaseval and \numpyeval. For example, as shown in Figure \ref{fig:figure1}, \texttt{pandas} is converted to {\texttt{monkey}}, \texttt{dataframe} is converted to \texttt{knowledgeframe}, and the API name \texttt{isin} is converted to \texttt{iscontain}. For more details on keyword conversion, please refer to Appendix~\ref{sec:keywords_dict_monkey_beat_num}. To craft the API documentation for Monkey and BeatNum, we manually paraphrased the descriptions of all the new APIs to ensure that the pre-trained language models have never seen them.

\section{Experiments}
In this section, we conduct experiments to illustrate the superiority of our proposed framework.

\subsection{Experimental Setup}
\paragraph{API Information.} 
\label{apiinfoasprompt}
As shown in the second column (\textit{APIs}) in Table~\ref{tab:main_results}, there are four settings for prompting API information before the context:
\begin{itemize}
    \item {\bf{No API}}: there is nothing to be prompted; 
    \item {\bf{Perfect}}: the information of golden APIs in the target code is prompted;
    \item {\bf{Top-$N$}}: the information of top $N$ APIs retrieved by \apiretriever is prompted, where $N\in\{1,2,3,5\}$;
    \item {\bf{Human}}: the information of the APIs chosen by users is prompted. In our experiments, we invited three volunteers who are programmers familiar with Python but without any background in our three benchmarks. As in Figure \ref{fig:human_in_the_loop_tests}, they interacted with the \apiretriever and provided their choices for all programming problems. The final APIs are determined by voting on their choices. 
\end{itemize}

\paragraph{Baselines.}
Our contributions can be reviewed in terms of both \apiretriever and \apicoder. For \apiretriever, all models in the No API setting are our baseline, while we propose the Perfect, Top-$N$, and Human settings. For \apicoder, the main baseline is our base model, \codegenmono $350$M~\cite{codegen}, in the same API information setting. We use \codegen for short in the following. In addition, we include advanced pre-trained code generation models that are comparable in parameter size: CodeT5~\cite{codet5}, CodeGPT~\cite{codegpt_codexglue}, CodeClippy\footnote{\url{https://github.com/CodedotAl/gpt-code-clippy}} and CodeParrot\footnote{\url{https://huggingface.co/lvwerra/codeparrot}}. Codex $12$B~\cite{codex} is also used to show the performance of giant models.

\begin{table*}
\centering
\resizebox{\linewidth}{!}{
\begin{tabular}{c|l|lll:lll:lll} 
\toprule
{\cellcolor[rgb]{0.831,0.831,0.831}}                                                                                                                                                  & \multicolumn{1}{c|}{\multirow{2}{*}{\begin{tabular}[c]{@{}c@{}}\textbf{\textbf{APIs}}\end{tabular}}} & \multicolumn{3}{c:}{\textbf{TorchDataEval}}                                                                                                               & \multicolumn{3}{c:}{\textbf{MonkeyEval}}                                                                                                    & \multicolumn{3}{c}{\textbf{\textbf{BeatNumEval}}}                                                                                                                   \\ 
\cdashline{3-11}
\multirow{-2}{*}{{\cellcolor[rgb]{0.831,0.831,0.831}}\textbf{APICoder}}                                                                                                               & \multicolumn{1}{c|}{}                                                                                                            & pass@$1$                                                & pass@$10$                                      & pass@$100$                                     & pass@$1$                                     & pass@$10$                                    & pass@$100$                                    & pass@$1$                                                & pass@$10$                                     & pass@$100$                                                \\ 
\hline
{\cellcolor[rgb]{0.831,0.831,0.831}}CodeT5 220M                                                                                                                                       & Top-2                                                                                                                            & 0.00                                                    & 0.00                                           & 0.00                                           & 0.00                                         & 0.00                                         & 0.00                                          & 0.00                                                    & 0.00                                          & 0.00                                                      \\ 
\hdashline
{\cellcolor[rgb]{0.831,0.831,0.831}}CodeGPT 124M                                                                                                                                      & Top-2                                                                                                                            & 0.67                                                    & 2.78                                           & 7.72                                           & 0.82                                         & 0.99                                         & 1.73                                          & 0.52                                                    & 1.88                                          & 4.70                                                      \\ 
\hdashline
{\cellcolor[rgb]{0.831,0.831,0.831}}CodeClippy 125M                                                                                                                                   & Top-2                                                                                                                            & 0.04                                                    & 0.39                                           & 2.75                                           & 0.10                                         & 0.76                                         & 1.86                                          & 0.03                                                    & 0.33                                          & 2.11                                                      \\ 
\hdashline
\rowcolor[rgb]{0.757,0.867,1} {\cellcolor[rgb]{0.831,0.831,0.831}}                                                                                                                    & No API                                                                                                                           & 4.04                                                    & 7.11                                           & 13.26                                          & 0.54                                         & 2.04                                         & 7.38                                          & 2.67                                                    & 7.66                                          & 18.86                                                     \\
\rowcolor[rgb]{1,0.925,0.792} {\cellcolor[rgb]{0.831,0.831,0.831}}                                                                                                                    & Perfect                                                                                                                          & 4.86\textsuperscript{\textcolor{red}{+0.82}}            & 8.88\textsuperscript{\textcolor{red}{+1.77}}   & 17.25\textsuperscript{\textcolor{red}{+3.99}}  & 2.39\textsuperscript{\textcolor{red}{+1.85}} & 3.33\textsuperscript{\textcolor{red}{+1.29}} & 9.99\textsuperscript{\textcolor{red}{+2.61}}  & 5.01\textsuperscript{\textcolor{red}{+2.34}}            & 11.30\textsuperscript{\textcolor{red}{+3.64}} & 26.36\textsuperscript{\textcolor{red}{+7.5}}              \\
{\cellcolor[rgb]{0.831,0.831,0.831}}                                                                                                                                                  & Top-1                                                                                                                            & 4.02\textsuperscript{\textcolor[rgb]{0,0.384,0}{-0.02}} & 8.35\textsuperscript{\textcolor{red}{+1.24}}   & 18.17\textsuperscript{\textcolor{red}{+4.91}}  & 2.54\textsuperscript{\textcolor{red}{+2.00}} & 3.43\textsuperscript{\textcolor{red}{+1.39}} & 11.39\textsuperscript{\textcolor{red}{+4.01}} & 4.32\textsuperscript{\textcolor{red}{+1.65}}            & 9.39\textsuperscript{\textcolor{red}{+1.73}}  & 19.91\textsuperscript{\textcolor{red}{+1.05}}             \\
{\cellcolor[rgb]{0.831,0.831,0.831}}                                                                                                                                                  & Top-2                                                                                                                            & 4.64\textsuperscript{\textcolor{red}{+0.60}}            & 8.96\textsuperscript{\textcolor{red}{+1.85}}   & 17.48\textsuperscript{\textcolor{red}{+4.22}}  & 1.52\textsuperscript{\textcolor{red}{+0.98}} & 2.96\textsuperscript{\textcolor{red}{+0.92}} & 9.32\textsuperscript{\textcolor{red}{+1.94}}  & 2.77\textsuperscript{\textcolor{red}{+0.10}}            & 8.57\textsuperscript{\textcolor{red}{+0.91}}  & 19.74\textsuperscript{\textcolor{red}{+0.88}}             \\
{\cellcolor[rgb]{0.831,0.831,0.831}}                                                                                                                                                  & Top-3                                                                                                                            & 4.00\textsuperscript{\textcolor[rgb]{0,0.384,0}{-0.04}} & 7.51\textsuperscript{\textcolor{red}{+0.40}}   & 15.13\textsuperscript{\textcolor{red}{+1.87}}  & 1.32\textsuperscript{\textcolor{red}{+0.78}} & 3.16\textsuperscript{\textcolor{red}{+1.12}} & 10.59\textsuperscript{\textcolor{red}{+3.21}} & 1.69\textsuperscript{\textcolor[rgb]{0,0.384,0}{-0.98}} & 9.01\textsuperscript{\textcolor{red}{+1.35}}  & 19.90\textsuperscript{\textcolor{red}{+1.04}}             \\
{\cellcolor[rgb]{0.831,0.831,0.831}}                                                                                                                                                  & Top-5                                                                                                                            & 4.22\textsuperscript{\textcolor{red}{+0.18}}            & 7.51\textsuperscript{\textcolor{red}{+0.40}}   & 15.43\textsuperscript{\textcolor{red}{+2.17}}  & 0.99\textsuperscript{\textcolor{red}{+0.45}} & 2.78\textsuperscript{\textcolor{red}{+0.74}} & 11.76\textsuperscript{\textcolor{red}{+4.38}} & 1.74\textsuperscript{\textcolor[rgb]{0,0.384,0}{-0.93}} & 8.11\textsuperscript{\textcolor{red}{+0.45}}  & 17.54\textsuperscript{\textcolor[rgb]{0,0.384,0}{-1.32}}  \\
\rowcolor[rgb]{0.933,0.859,0.933} \multirow{-7}{*}{{\cellcolor[rgb]{0.831,0.831,0.831}}\begin{tabular}[c]{@{}>{\cellcolor[rgb]{0.831,0.831,0.831}}c@{}}CodeParrot\\110M\end{tabular}} & Human                                                                                                                            & 4.01\textsuperscript{\textcolor[rgb]{0,0.384,0}{-0.03}} & 7.60\textsuperscript{\textcolor{red}{+0.49}}   & 14.47\textsuperscript{\textcolor{red}{+1.21}}  & 2.44\textsuperscript{\textcolor{red}{+1.90}} & 3.62\textsuperscript{\textcolor{red}{+1.58}} & 9.83\textsuperscript{\textcolor{red}{+2.45}}  & 5.23\textsuperscript{\textcolor{red}{+2.56}}            & 11.78\textsuperscript{\textcolor{red}{+4.12}} & 22.81\textsuperscript{\textcolor{red}{+3.95}}             \\ 
\hdashline
\rowcolor[rgb]{0.757,0.867,1} {\cellcolor[rgb]{0.831,0.831,0.831}}                                                                                                                    & No API                                                                                                                           & 6.72                                                    & 15.71                                          & 22.00                                          & 0.95                                         & 4.90                                         & 8.89                                          & 5.15                                                    & 11.96                                         & 18.79                                                     \\
\rowcolor[rgb]{1,0.925,0.792} {\cellcolor[rgb]{0.831,0.831,0.831}}                                                                                                                    & Perfect                                                                                                                          & 9.84\textsuperscript{\textcolor{red}{+3.12}}            & 22.62\textsuperscript{\textcolor{red}{+6.91}}  & 34.00\textsuperscript{\textcolor{red}{+12.00}} & 2.14\textsuperscript{\textcolor{red}{+1.19}} & 6.41\textsuperscript{\textcolor{red}{+1.51}} & 11.86\textsuperscript{\textcolor{red}{+2.97}} & 9.47\textsuperscript{\textcolor{red}{+4.32}}            & 17.05\textsuperscript{\textcolor{red}{+5.09}} & 28.67\textsuperscript{\textcolor{red}{+9.88}}             \\
{\cellcolor[rgb]{0.831,0.831,0.831}}                                                                                                                                                  & Top-1                                                                                                                            & 8.72\textsuperscript{\textcolor{red}{+2.00}}            & 19.22\textsuperscript{\textcolor{red}{+3.51}}  & 27.97\textsuperscript{\textcolor{red}{+5.97}}  & 2.22\textsuperscript{\textcolor{red}{+1.27}} & 7.20\textsuperscript{\textcolor{red}{+2.30}} & 12.85\textsuperscript{\textcolor{red}{+3.96}} & 7.52\textsuperscript{\textcolor{red}{+2.37}}            & 15.25\textsuperscript{\textcolor{red}{+3.29}} & 24.71\textsuperscript{\textcolor{red}{+5.92}}             \\
{\cellcolor[rgb]{0.831,0.831,0.831}}                                                                                                                                                  & Top-2                                                                                                                            & 7.52\textsuperscript{\textcolor{red}{+0.80}}            & 16.36\textsuperscript{\textcolor{red}{+0.65}}  & 26.00\textsuperscript{\textcolor{red}{+4.00}}  & 2.46\textsuperscript{\textcolor{red}{+1.51}} & 6.35\textsuperscript{\textcolor{red}{+1.45}} & 9.89\textsuperscript{\textcolor{red}{+1.00}}  & 6.65\textsuperscript{\textcolor{red}{+1.50}}            & 13.68\textsuperscript{\textcolor{red}{+1.72}} & 22.74\textsuperscript{\textcolor{red}{+3.95}}             \\
{\cellcolor[rgb]{0.831,0.831,0.831}}                                                                                                                                                  & Top-3                                                                                                                            & 7.92\textsuperscript{\textcolor{red}{+1.20}}            & 18.65\textsuperscript{\textcolor{red}{+2.94}}  & 28.00\textsuperscript{\textcolor{red}{+6.00}}  & 2.02\textsuperscript{\textcolor{red}{+1.07}} & 5.26\textsuperscript{\textcolor{red}{+0.36}} & 8.89\textsuperscript{\textcolor{red}{+0.00}}  & 6.26\textsuperscript{\textcolor{red}{+1.11}}            & 16.12\textsuperscript{\textcolor{red}{+4.16}} & 24.72\textsuperscript{\textcolor{red}{+5.93}}             \\
{\cellcolor[rgb]{0.831,0.831,0.831}}                                                                                                                                                  & Top-5                                                                                                                            & 6.08\textsuperscript{\textcolor[rgb]{0,0.384,0}{-0.64}} & 17.48\textsuperscript{\textcolor{red}{+1.77}}  & 25.95\textsuperscript{\textcolor{red}{+3.95}}  & 1.58\textsuperscript{\textcolor{red}{+0.63}} & 5.45\textsuperscript{\textcolor{red}{+0.55}} & 9.88\textsuperscript{\textcolor{red}{+0.99}}  & 6.34\textsuperscript{\textcolor{red}{+1.19}}            & 15.05\textsuperscript{\textcolor{red}{+3.09}} & 21.76\textsuperscript{\textcolor{red}{+2.97}}             \\
\rowcolor[rgb]{0.933,0.859,0.933} \multirow{-7}{*}{{\cellcolor[rgb]{0.831,0.831,0.831}}\begin{tabular}[c]{@{}>{\cellcolor[rgb]{0.831,0.831,0.831}}c@{}}\codegen\\350M\end{tabular}}    & Human                                                                                                                            & 8.08\textsuperscript{\textcolor{red}{+1.36}}            & 19.85\textsuperscript{\textcolor{red}{+4.14}}  & 31.95\textsuperscript{\textcolor{red}{+9.95}}  & 2.14\textsuperscript{\textcolor{red}{+1.19}} & 6.14\textsuperscript{\textcolor{red}{+1.24}} & 11.86\textsuperscript{\textcolor{red}{+2.97}} & 9.47\textsuperscript{\textcolor{red}{+4.32}}            & 17.12\textsuperscript{\textcolor{red}{+5.06}} & 28.67\textsuperscript{\textcolor{red}{+9.88}}             \\ 
\hdashline
\rowcolor[rgb]{0.757,0.867,1} {\cellcolor[rgb]{0.831,0.831,0.831}}                                                                                                                    & No API                                                                                                                           & 7.19                                                    & 16.93                                          & 23.97                                          & 1.19                                         & 4.68                                         & 7.91                                          & 4.44                                                    & 8.24                                          & 13.83                                                     \\
\rowcolor[rgb]{1,0.925,0.792} {\cellcolor[rgb]{0.831,0.831,0.831}}                                                                                                                    & Perfect                                                                                                                          & 20.23\textsuperscript{\textcolor{red}{+13.04}}          & 33.37\textsuperscript{\textcolor{red}{+16.44}} & 41.97\textsuperscript{\textcolor{red}{+18.00}} & 4.59\textsuperscript{\textcolor{red}{+3.40}} & 9.14\textsuperscript{\textcolor{red}{+4.46}} & 13.85\textsuperscript{\textcolor{red}{+5.94}} & 9.62\textsuperscript{\textcolor{red}{+5.18}}            & 16.51\textsuperscript{\textcolor{red}{+8.27}} & 22.75\textsuperscript{\textcolor{red}{+8.92}}             \\
{\cellcolor[rgb]{0.831,0.831,0.831}}                                                                                                                                                  & Top-1                                                                                                                            & 12.89\textsuperscript{\textcolor{red}{+5.70}}           & 24.26\textsuperscript{\textcolor{red}{+7.33}}  & 31.97\textsuperscript{\textcolor{red}{+8.00}}  & 2.89\textsuperscript{\textcolor{red}{+1.70}} & 8.28\textsuperscript{\textcolor{red}{+3.60}} & 12.86\textsuperscript{\textcolor{red}{+4.94}} & 6.61\textsuperscript{\textcolor{red}{+2.17}}            & 12.62\textsuperscript{\textcolor{red}{+4.38}} & 17.80\textsuperscript{\textcolor{red}{+3.97}}             \\
{\cellcolor[rgb]{0.831,0.831,0.831}}                                                                                                                                                  & Top-2                                                                                                                            & 10.41\textsuperscript{\textcolor{red}{+3.22}}           & 23.50\textsuperscript{\textcolor{red}{+6.57}}  & 31.98\textsuperscript{\textcolor{red}{+8.01}}  & 3.41\textsuperscript{\textcolor{red}{+2.22}} & 8.33\textsuperscript{\textcolor{red}{+3.65}} & 11.87\textsuperscript{\textcolor{red}{+8.90}} & 5.90\textsuperscript{\textcolor{red}{+1.46}}            & 11.79\textsuperscript{\textcolor{red}{+3.55}} & 15.83\textsuperscript{\textcolor{red}{+2.00}}             \\
{\cellcolor[rgb]{0.831,0.831,0.831}}                                                                                                                                                  & Top-3                                                                                                                            & 10.49\textsuperscript{\textcolor{red}{+3.30}}           & 25.45\textsuperscript{\textcolor{red}{+8.52}}  & 35.98\textsuperscript{\textcolor{red}{+12.01}} & 3.17\textsuperscript{\textcolor{red}{+1.98}} & 7.51\textsuperscript{\textcolor{red}{+2.83}} & 10.88\textsuperscript{\textcolor{red}{+2.97}} & 5.11\textsuperscript{\textcolor{red}{+0.67}}            & 11.40\textsuperscript{\textcolor{red}{+3.16}} & 15.82\textsuperscript{\textcolor{red}{+1.99}}             \\
{\cellcolor[rgb]{0.831,0.831,0.831}}                                                                                                                                                  & Top-5                                                                                                                            & 10.34\textsuperscript{\textcolor{red}{+3.15}}           & 23.04\textsuperscript{\textcolor{red}{+6.11}}  & 27.99\textsuperscript{\textcolor{red}{+4.02}}  & 1.94\textsuperscript{\textcolor{red}{+0.75}} & 4.75\textsuperscript{\textcolor{red}{+0.07}} & 7.91\textsuperscript{\textcolor{red}{+0.00}}  & 5.07\textsuperscript{\textcolor{red}{+0.63}}            & 9.64\textsuperscript{\textcolor{red}{+1.40}}  & 13.84\textsuperscript{\textcolor{red}{+0.01}}             \\
\rowcolor[rgb]{0.933,0.859,0.933} \multirow{-7}{*}{{\cellcolor[rgb]{0.831,0.831,0.831}}\begin{tabular}[c]{@{}>{\cellcolor[rgb]{0.831,0.831,0.831}}c@{}}\codegenapi\\350M\end{tabular}} & Human                                                                                                                            & 15.57\textsuperscript{\textcolor{red}{+8.38}}           & 27.76\textsuperscript{\textcolor{red}{+10.83}} & 33.97\textsuperscript{\textcolor{red}{+10.00}} & 3.76\textsuperscript{\textcolor{red}{+2.57}} & 8.32\textsuperscript{\textcolor{red}{+3.64}} & 12.86\textsuperscript{\textcolor{red}{+4.95}} & 9.39\textsuperscript{\textcolor{red}{+4.95}}            & 16.40\textsuperscript{\textcolor{red}{+8.16}} & 23.74\textsuperscript{\textcolor{red}{+9.91}}             \\ 
\hline\hline
\rowcolor[rgb]{0.757,0.867,1} {\cellcolor[rgb]{0.831,0.831,0.831}}                                                                                                                    & No API                                                                                                                           & 7.16                                                    & 14.46                                          & 23.75                                          & 1.47                                         & 3.53                                         & 7.31                                          & 6.95                                                    & 17.54                                         & 25.57                                                     \\
\rowcolor[rgb]{1,0.925,0.792} {\cellcolor[rgb]{0.831,0.831,0.831}}                                                                                                                    & Perfect                                                                                                                          & 25.03\textsuperscript{\textcolor{red}{+17.87}}          & 51.26\textsuperscript{\textcolor{red}{+36.80}} & 56.75\textsuperscript{\textcolor{red}{+33.00}} & 3.58\textsuperscript{\textcolor{red}{+2.11}} & 7.48\textsuperscript{\textcolor{red}{+3.95}} & 12.61\textsuperscript{\textcolor{red}{+5.30}} & 8.59\textsuperscript{\textcolor{red}{+1.64}}            & 23.75\textsuperscript{\textcolor{red}{+6.21}} & 36.99\textsuperscript{\textcolor{red}{+11.42}}            \\
\multirow{-3}{*}{{\cellcolor[rgb]{0.831,0.831,0.831}}\begin{tabular}[c]{@{}>{\cellcolor[rgb]{0.831,0.831,0.831}}c@{}}Codex\\12B\end{tabular}}                                         & Top-2~                                                                                                                           & 17.98\textsuperscript{\textcolor{red}{+10.82}}          & 32.75\textsuperscript{\textcolor{red}{+18.29}} & 41.51\textsuperscript{\textcolor{red}{+17.76}} & 1.92\textsuperscript{\textcolor{red}{+0.45}} & 5.91\textsuperscript{\textcolor{red}{+2.38}} & 11.08\textsuperscript{\textcolor{red}{+3.77}} & 9.54\textsuperscript{\textcolor{red}{+2.59}}            & 21.77\textsuperscript{\textcolor{red}{+4.23}} & 32.45\textsuperscript{\textcolor{red}{+6.88}}             \\
\bottomrule
\end{tabular}
}
\caption{Pass$@k$($\%$) results on the three benchmarks. The \textcolor[rgb]{0.325,0.592,1}{blue} background means no API as extra prompt; the \textcolor[rgb]{1,0.647,0}{yellow} background means perfect APIs as extra prompt; the write background means top-$1,2,3,$ or $5$ APIs retrieved by \apiretriever as extra prompt; and the \textcolor[rgb]{0.502,0,0.502}{purple} background means the APIs chosen by human from top-$5$ of \apiretriever as extra prompt. Numbers in \textcolor{red}{red} and \textcolor[rgb]{0,0.392,0}{green} indicate the absolute changes over no API setting.}
\label{tab:main_results}
\end{table*}

\paragraph{Evaluation Metrics.}
Followed by \citet{codex}, we regard pass$@k$ as our metric. For each programming problem, we sample $n\ge k$ code snippets, and then count the number of correct ones $c$, where passing all test cases is considered as correct. If $n-c<k$, then pass$@k$ equals $1$; otherwise, equals $1-\prod\nolimits_{i=n-c+1}^{n} (1-\frac{k}{i})$. In our experiments, $k$ is set to one of $[1, 10, 100]$ and $n$ is set to $200$.

\paragraph{Implementation Details.}
We implement our approach based on PyTorch~\cite{pytorch} and Huggingface's transformers~\cite{transformers}. We use a dense retrieval toolkit\footnote{\url{https://github.com/luyug/Dense}} to train \apiretriever by setting the batch size to $10$ per device, the learning rate to $1$e-$5$, the ratio of positive vs. negative samples to $1$:$8$, and the vector dimensions $z$ of $\mathbf{p}$ and $\mathbf{a}$ to $768$. The model uses cross-entropy as the loss function and Adam~\cite{adam} as the parameters optimizer. It is trained for $100$K steps on a cluster of $8$ NVIDIA V$100$ GPUs with $32$GB memory. The training time is about $3$ days. For pre-training \codegenapi, we set the code block size to $1,024$, the batch size to $4$, the learning rate to $5$e-$4$, the gradient accumulation steps to $4$, the weight decay to $0.1$, and the warm up steps to $1,000$. Noise APIs are added at a rate of $0.05$. It is trained for $100$K steps about $1.6$ days on $8$ $32$GB NVIDIA V$100$ GPUs. In all of our training phases, we use mixed precision FP$16$ to speed up. When generating code snippets using pre-trained models, we conduct various temperatures ranging from $0.1$ to $1.0$ with the interval of $0.1$. All results are reported with the best values across these hyper-parameters.

\subsection{Main Results} \label{main_results}
Table~\ref{tab:main_results} summarizes the performance of our framework and all baselines on \torchdataeval, \monkeyeval, and \beatnumeval. Based on numerous experimental results, we derived plausible observations and valuable insights to answer the following research questions.

\emph{``Is API information useful for private library oriented code generation?''} 
As we can see in Table~\ref{tab:main_results}, all models without prompting any APIs (the No API setting) achieve relatively poor performance on all benchmarks. Especially, Codex $12$B, a powerful code generation model with large parameters, can only achieve similar performance to \codegen and \codegenapi $350$M in the No API setting. This indicates that even with gigantic models, the task of code generation with private libraries is extremely challenging. Encouragingly, with prompted API information (the Perfect, Top-$N$, Human settings), both the off-the-shelf models (e.g., CodeParrot, \codegen, and Codex) and our continually pre-trained CodeGenAPI achieve consistent performance gains compared to those in the No API setting. Moreover, the more powerful the model itself in code generation (i.e., Codex $12$B > \codegen $350$M > CodeParrot $110$M), the more benefits that API information can bring. For example, on \torchdataeval in the Perfect setting, Codex $12$B brings pass@$10$ an absolute improvement of $36.89\%$, while CodeParrot $110$M only improves $1.77\%$. This observation also suggests that prompting API information can unleash the potential of gigantic models towards invoking private APIs. All the above results prove the usefulness of API information for code generation for private libraries. 

\emph{``Is the \apiretriever effective in finding useful API information?''} 
All models in the Top-$N$ setting outperform the same models in the No API setting, suggesting that \apiretriever is able to find useful APIs. For a certain model, we observe that the Top-$1$/Top-$2$ settings usually perform better than the Top-$3$/Top-$5$ settings due to the fact that the latter introduces more noise APIs to the \apicoder. In addition, involving humans (the Human setting) in the selection of APIs can further improve performance, suggesting the effectiveness of the human interaction we designed. Note that the Top-$N$ and Human settings are occasionally superior to the Perfect setting, which is reasonable because the noise APIs exist when training the model.

\begin{figure}[t]
    \small
    \centering
    \includegraphics[width=1\linewidth]{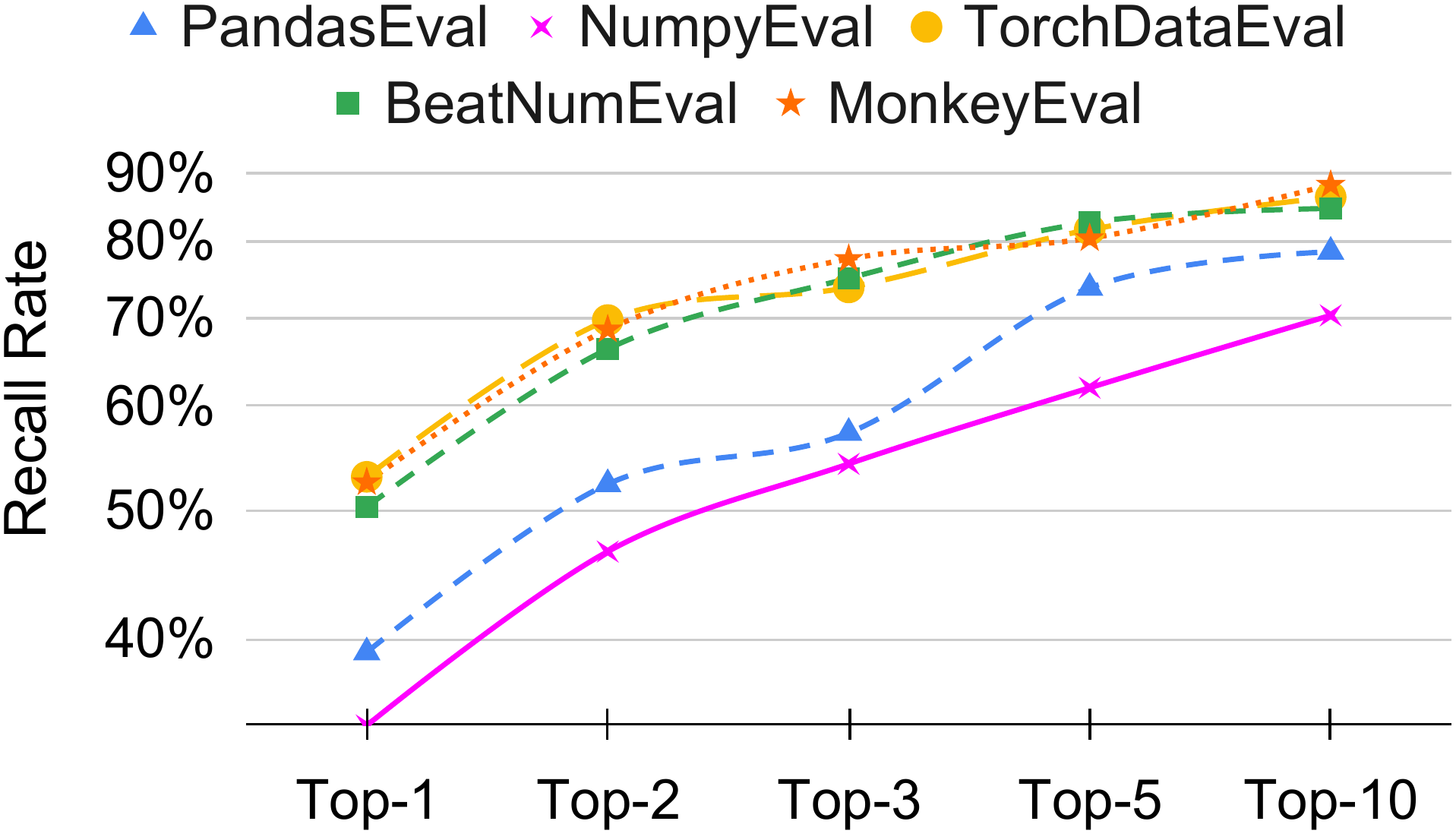}
    \caption{The recall rates of retrieved APIs.}% on five benchmarks.}
    \label{fig:recall_rate}
\end{figure}

\emph{``Is the \apicoder effective in invoking private APIs?''}
As shown in Table~\ref{tab:main_results}, off-the-shelf models like \codegen are capable of handling private library invocations. To seek more extraordinary performance, we continually pre-train \codegen and obtain a new model \codegenapi. We can observe that \codegenapi consistently outperforms its base model \codegen on \torchdataeval and \monkeyeval, which proves the effectiveness of \codegenapi. However, on \beatnumeval, \codegenapi is inferior to \codegen. After careful troubleshooting, we reveal that the process of continual pre-training aims to essentially learn how to invoke the correct APIs with maximum likelihood, while the key obstacle to using BeatNum modified from Numpy lies in the numerical calculation like `\texttt{a[:,None]+b*2}', instead of invoking the correct APIs. Therefore, \codegenapi fails to yield benefits for \beatnumeval. Overall, \apicoder has the capability to invoke private APIs.

\begin{table*}
\centering
\resizebox{\linewidth}{!}{
\begin{tabular}{l|ccc:ccc:ccc} 
\toprule
\multirow{2}{*}{\textbf{APICoder}} & \multicolumn{3}{c:}{\textbf{\textbf{TorchDataEval}}} & \multicolumn{3}{c:}{\textbf{MonkeyEval}}       & \multicolumn{3}{c}{\textbf{\textbf{BeatNumEval}}}  \\ 
\cdashline{2-10}
                                   & pass@1         & pass@10        & pass@100           & pass@1        & pass@10       & pass@100       & pass@1        & pass@10        & pass@100          \\ 
\hline
\codegenapi                         & \textbf{10.41} & \textbf{23.50} & \textbf{31.98}     & \textbf{3.41} & \textbf{8.33} & \textbf{11.87} & \textbf{5.90} & \textbf{11.79} & \textbf{15.83}    \\ 
\hdashline
~ -w/~noise~rate $0$\%             & 9.41           & 22.88          & 31.08              & 2.69          & 8.03          & 11.18          & 5.77          & 11.01          & 14.52             \\
~~-w/~noise~rate $10$\%            & 9.19           & 22.87          & 30.98              & 3.04          & 7.67          & 11.10          & 4.99          & 10.80          & 15.18             \\
~ -w/~noise~rate $20$\%            & 8.92           & 23.04          & 30.57              & 2.00          & 7.39          & 10.64          & 4.48          & 10.97          & 13.41             \\
~ -w/o resampling                  & 8.65           & 21.00          & 29.71              & 2.47          & 7.96          & 10.13          & 5.21          & 8.68           & 14.75             \\
\bottomrule
\end{tabular}
}
\caption{Ablation studies for \codegenapi in the Top-2 setting (top 2 APIs provided by \apiretriever are prompted). The default setting of \codegenapi is to use the resampling strategy and a noise rate of $5$\%.}
\label{tab:ablation}
\end{table*}

\subsection{Closer Analysis}
We have demonstrated the effectiveness of our framework. In this subsection, we provide several closer analyses to inspire future work in this direction. 

\begin{figure}[t]
    \small
    \centering
    \includegraphics[width=0.93\linewidth]{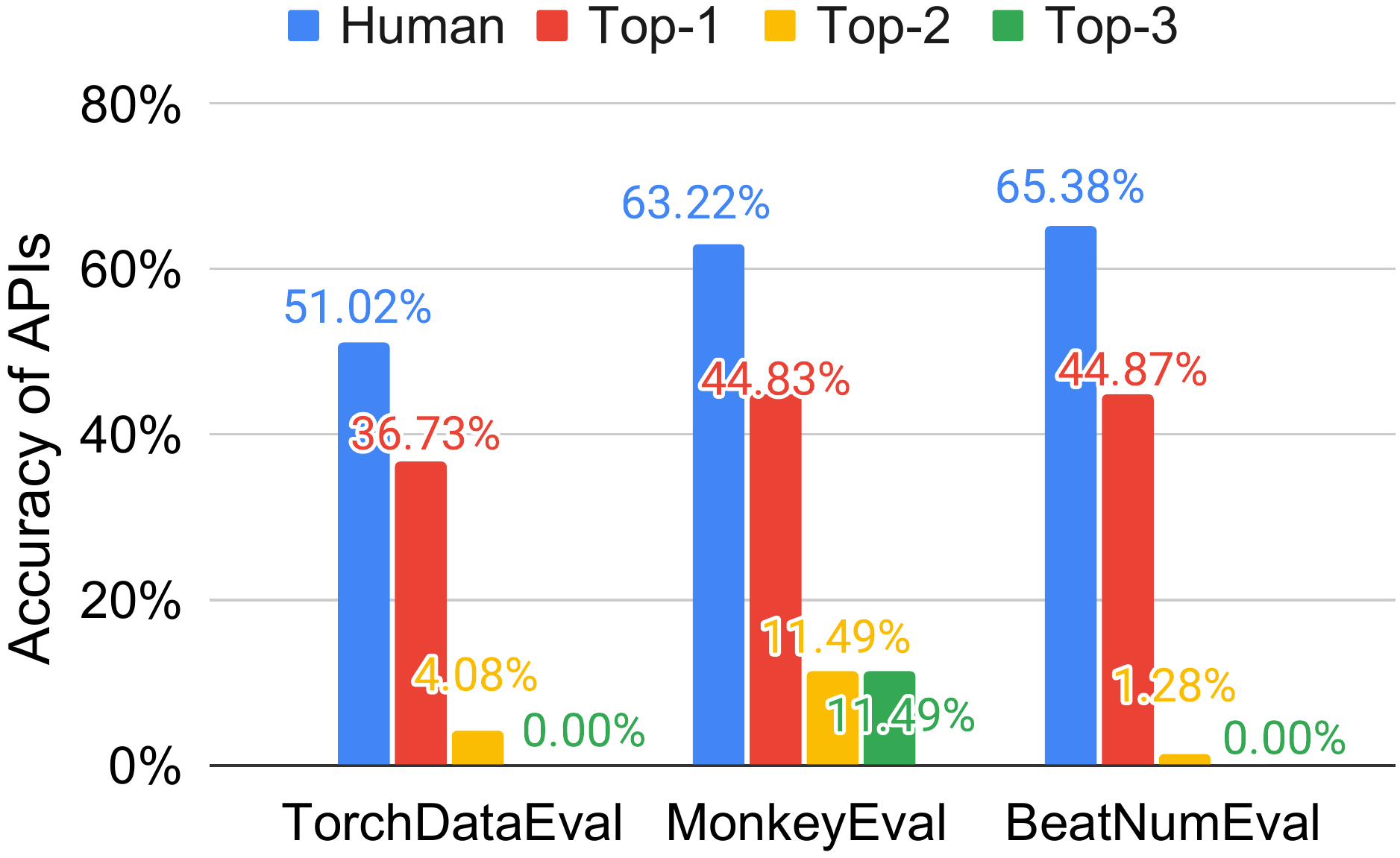}
    \caption{Accuracy of retrieved APIs.}
    \label{fig:api_acc}
\end{figure}

\paragraph{Quality of Retrieved APIs.}
Retrieving the correct APIs as prompts can enhance the code generation performance for private libraries, so we would like to evaluate the effectiveness of \apiretriever.  Figure~\ref{fig:recall_rate} shows the recall rate of \apiretriever on five benchmarks. We can see that the recall rates of top-$5$ are already high, demonstrating that it is reasonable to provide $5$ API candidates for users to choose from. Furthermore, as shown in Figure~\ref{fig:api_acc}, we analyze the accuracy of APIs chosen by users. We observe that it dramatically exceeds the accuracy of top $1,2$ or $3$ APIs retrieved by \apiretriever. This suggests that it is feasible to involve humans in the retrieval of APIs.

\begin{figure}[t]
    \small
    \centering
    \includegraphics[width=1\linewidth]{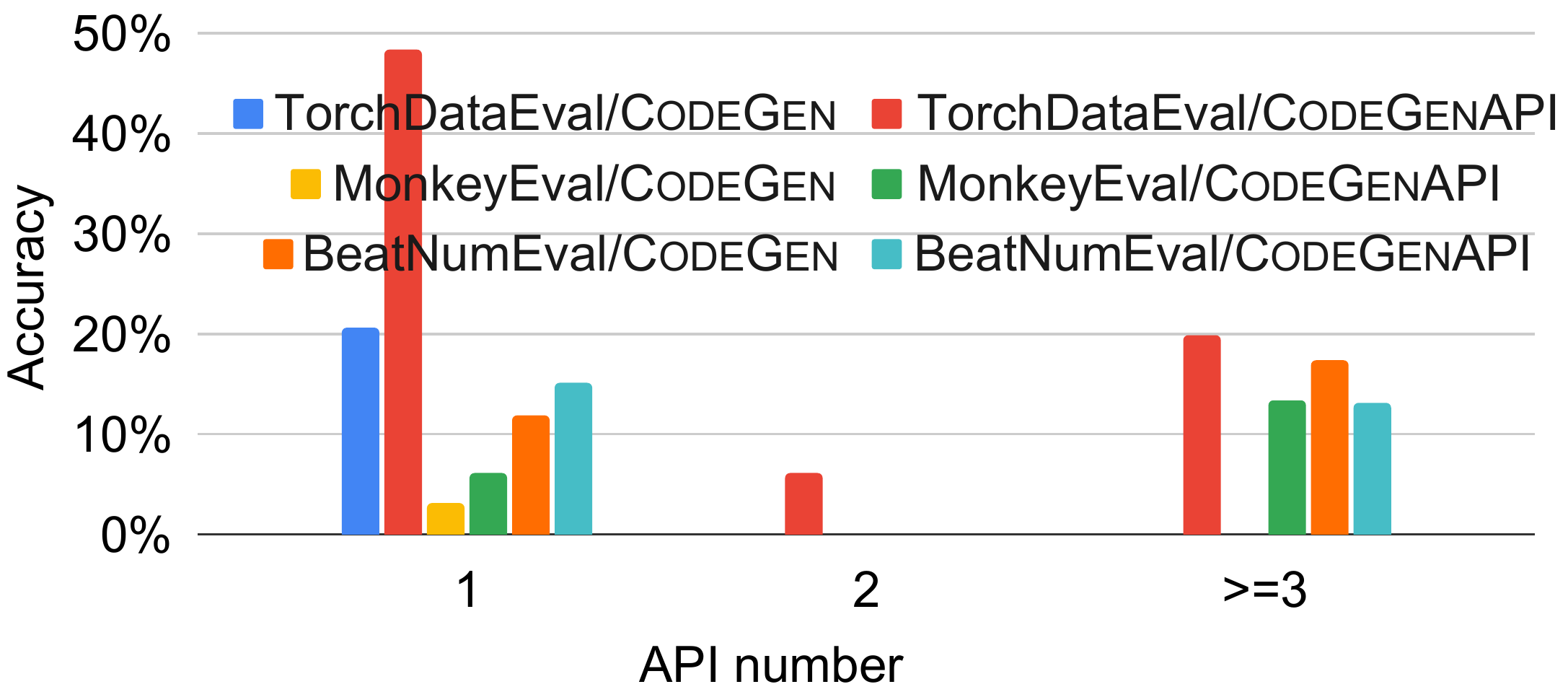}
    \caption{Accuracy of \codegenapi and \codegen with respect to the number of APIs. The problem is solved if one of $200$ samples passes all test cases.}
    \label{fig:api_number}
\end{figure}

\paragraph{Different Difficulty.}
We would like to explore the performance of \codegenapi on varying difficulty problems. So we calculate its accuracy across various numbers of APIs in target code $\mathbf{y}$. Each benchmark is divided into $3$ parts, according to the number of APIs. Figure~\ref{fig:api_number} shows that \codegenapi outperforms \codegen by a large margin on the problems containing only one API. The trend still holds as the number of APIs increases. It demonstrates \codegenapi can boost the performance of generating code snippets using private library on varying difficulty. 

\paragraph{Noise Rate.}
A well-chosen noise rate can improve the robustness of \codegenapi against a variety of APIs. If we set the noise rate too large, it may change the original distribution of the code corpus, while too small will lose the capability to deal with noise APIs. The default noise rate is $5$\%, and we also try $0$\%, $10$\%, and $20$\%. As shown in Table~\ref{tab:ablation}, both too large and too small noise rates can degrade the performance.

\paragraph{Resampling Strategy.}
Making high-quality python files high-priority, and vice versa, is in line with our intuition. To demonstrate it, we remove the resampling strategy as mentioned in Section~\ref{sec:apicoder}. As shown in Table~\ref{tab:ablation}, we observe a steady decline in performance on the three benchmarks. Such an observation demonstrates the effectiveness of the sampling strategy.

\paragraph{\codegenapi for Public Library.} \label{codegenapiforpub}
Technically speaking, \codegenapi also can be employed for generating code for public libraries. So, we do experiments on \pandaseval and \numpyeval and show the results in Table~\ref{tab:pandas_numpy}. We find that the performance improvement of \codegenapi over the base model on public libraries is not as significant as on private libraries. One major reason is that the models have seen the public libraries during pre-training, so prompting API information yields limited benefit. We can see \codegenapi excels over \codegen when prompting perfect APIs. But when prompting top-2 APIs, the advantages of \codegenapi are not exhibited. This means that \codegenapi can also work on third-party public libraries, but it depends heavily on the performance of \apiretriever.

\begin{table}
\centering
\resizebox{70mm}{!}{
\begin{tabular}{c|l|ccc} 
\toprule
\multirow{2}{*}{\textbf{APICoder}} & \multicolumn{1}{c|}{\multirow{2}{*}{\begin{tabular}[c]{@{}c@{}}\textbf{APIs}\end{tabular}}} & \multicolumn{3}{c}{\textbf{PandasEval}}                                                                                \\ 
\cdashline{3-5}
                                   & \multicolumn{1}{c|}{}                                                                                          & pass@1                                & pass@10                               & pass@100                               \\ 
\hline
\multirow{3}{*}{\codegen}           & {\cellcolor[rgb]{0.757,0.867,1}}No API                                                                         & {\cellcolor[rgb]{0.757,0.867,1}}14.24 & {\cellcolor[rgb]{0.757,0.867,1}}30.71 & {\cellcolor[rgb]{0.757,0.867,1}}46.04  \\
                                   & {\cellcolor[rgb]{1,0.925,0.792}}Perfect                                                                        & {\cellcolor[rgb]{1,0.925,0.792}}11.21 & {\cellcolor[rgb]{1,0.925,0.792}}33.59 & {\cellcolor[rgb]{1,0.925,0.792}}48.47  \\
                                   & Top-2                                                                                                          & 9.54                                  & 29.02                                 & 40.56                                  \\ 
\hdashline
\multirow{3}{*}{\codegenapi}        & {\cellcolor[rgb]{0.757,0.867,1}}No API                                                                         & {\cellcolor[rgb]{0.757,0.867,1}}13.58 & {\cellcolor[rgb]{0.757,0.867,1}}34.95 & {\cellcolor[rgb]{0.757,0.867,1}}46.51  \\
                                   & {\cellcolor[rgb]{1,0.925,0.792}}Perfect                                                                        & {\cellcolor[rgb]{1,0.925,0.792}}19.96 & {\cellcolor[rgb]{1,0.925,0.792}}42.36 & {\cellcolor[rgb]{1,0.925,0.792}}53.43  \\
                                   & Top-2                                                                                                          & 11.25                                 & 28.61                                 & 39.48                                  \\ 
\hdashline
                                   &                                                                                                                & \multicolumn{3}{c}{\textbf{NumpyEval}}                                                                                 \\ 
\hdashline
\multirow{3}{*}{\codegen}           & {\cellcolor[rgb]{0.757,0.867,1}}No API                                                                         & {\cellcolor[rgb]{0.757,0.867,1}}19.31 & {\cellcolor[rgb]{0.757,0.867,1}}40.89 & {\cellcolor[rgb]{0.757,0.867,1}}60.58  \\
                                   & {\cellcolor[rgb]{1,0.925,0.792}}Perfect                                                                        & {\cellcolor[rgb]{1,0.925,0.792}}21.41 & {\cellcolor[rgb]{1,0.925,0.792}}41.08 & {\cellcolor[rgb]{1,0.925,0.792}}56.38  \\
                                   & Top-2                                                                                                          & 18.30                                 & 35.12                                 & 48.46                                  \\ 
\hdashline
\multirow{3}{*}{\codegenapi}        & {\cellcolor[rgb]{0.757,0.867,1}}No API                                                                         & {\cellcolor[rgb]{0.757,0.867,1}}16.55 & {\cellcolor[rgb]{0.757,0.867,1}}29.48 & {\cellcolor[rgb]{0.757,0.867,1}}42.52  \\
                                   & {\cellcolor[rgb]{1,0.925,0.792}}Perfect                                                                        & {\cellcolor[rgb]{1,0.925,0.792}}24.83 & {\cellcolor[rgb]{1,0.925,0.792}}41.47 & {\cellcolor[rgb]{1,0.925,0.792}}54.41  \\
                                   & Top-2                                                                                                          & 12.67                                 & 27.32                                 & 35.62                                  \\
\bottomrule
\end{tabular}
}
\caption{Results of \codegen and \codegenapi on \pandaseval and \numpyeval.}
\label{tab:pandas_numpy}
\end{table}

\section{Related Work} \label{sec:related_work}
\subsection{Code Generation} \label{sec:rw_code_generation}
Thanks to the recent development of pre-training techniques~\cite{bert,T5,BART,yan2021large,gpt3}, a lot of pre-trained language models have been proposed for code-related tasks. For example, CuBERT~\cite{cubert}, CodeBERT~\cite{codebert}, GraphCodeBERT~\cite{graphcodebert}, CodeT5~\cite{codet5}, CodeGPT~\cite{codegpt_codexglue}, PLBART~\cite{plbart}, PyCodeGPT~\cite{cert}, \codegen~\cite{codegen}, Codex~\cite{codex}, AlphaCode~\cite{alphacode}, and InCoder~\cite{incoder}. Almost all of them focus on standalone code, while JigSaw~\cite{jigsaw} and CERT~\cite{cert} are presented for generating code using public libraries. In this paper, we aim to generate code invoking private APIs, which is a common scenario in practice. It is more challenging because pre-trained language models have never seen any information about private libraries. As for benchmarks, HumanEval~\cite{codex}, APPs~\cite{apps}, P3~\cite{p3}, MBPP~\cite{mbpp}, BIG-bench~\cite{bigbench}, and CodeContests~\cite{alphacode} were proposed to evaluate the performance of generating standalone code. GSM8K-Python~\cite{GSM8K-Python} and MathQA-Python~\cite{mbpp} were engaged in evaluating the capability of solving mathematical problems. \pandaseval and \numpyeval~\cite{cert} were released to evaluate the code generation for public library. We propose three benchmarks, called \torchdataeval, \monkeyeval, and \beatnumeval, aiming to evaluate the performance of code generation for private library.

\subsection{Retrieval-Based Generation} \label{sec:rw_retrieval_based_generation}
In the natural language field, retrieval-based generation is a hot topic. A lot of works~\cite{fusion_in_decoder,dense-retrieval,rocketqa,xiong2020approximate,santhanam2021colbertv2,s2ql,formal2022distillation} have emerged under this topic. Therefore, we refer to the above methods and design our \apiretriever for private API retrieval. 
In the programming language field, there are also several attempts to use retrieval techniques, such as \textsc{DeepAPI}~\cite{deep_api_learning}, REDCODER~\cite{redcoder}, ReACC~\cite{reacc}, and DocCoder~\cite{doccoder}. 
Our work is fundamentally different from them. They all aim to retrieve public code snippets or other resources on GitHub/StackOverFlow based on the user query, while our goal is to retrieve APIs from the API documentation of private library based on code comments. Besides, we design retrieval because we focus on private APIs, which have not been seen by the pre-trained generative language models.

\section{Conclusion}
In this paper, we propose a novel framework for code generation for private library. There are two modules: for a specific programming problem, \apiretriever first finds out the useful private APIs from API documentation, and then \apicoder leverages these APIs to generate the code. We craft three benchmarks, including \torchdataeval, \monkeyeval, and \beatnumeval, for better evaluating private library oriented code generation. The experimental results and thorough analysis demonstrate the reasonableness and effectiveness of our framework. In future work, we would like to explore how to make better use of API documentation for code generation and improve the approach for real use when programming with private libraries.

\section*{Limitations}
While our proposed approach exhibits many advantages, it also has a few limitations. 
(1) As stated in Section~\ref{main_results}, our approach that prompts APIs for programming problem relies heavily upon the code generation capacity of the language model itself. The more powerful the model itself, the more benefits the prompting APIs bring. Likewise, we also find that if a model itself shows very poor performance, prompting APIs will not bring any benefit to it or even bring negative effects. 
(2) As the first navigator to explore code generation with private library, we have built three private libraries, but they all include a relatively small number of APIs (<$200$). With these APIs, our \apiretriever can exhibit decent performance. But we surmise that it may become more challenging for \apiretriever as the number of APIs increases. 
(3) It is extremely challenging to find a real private library and craft a benchmark like \torchdataeval. To evaluate our idea quickly and cost-effectively, besides \torchdataeval, we also crafted two pseudo private libraries that are modified from the existing public ones as mentioned in Section~\ref{sec:benchmark}. Although we have done our best to preserve the two pseudo private libraries in line with the real private library, it may still pose some threats to the fair evaluation of code generation for private library. 
(4) We can see from Table~\ref{tab:main_results} that most models with the Top-N setting fall behind the same model with the Perfect setting. Such observation demonstrates that \apiretriever we designed has a big room for improvement.
(5) Our experiments show that our framework can enhance the quality of private library oriented code generation on Python. Limitations may exist when we generalize it to other programming languages such as Java, C, and C++ since the characteristics of libraries for different programming languages are slightly different. 

\bibliography{anthology,custom}
\bibliographystyle{acl_natbib}
\appendix

\section{Collection of API Documentation} \label{sec:31lib}
We aim to use data from public libraries for training and generalize the models to private libraries. Thus, we crawled the API documentation of the $31$ most popular public libraries in Python. Table~\ref{tab:31libs} summarizes the number of APIs we extracted for each library.

\section{Resampling Strategy} \label{appendix:resampling_strategy}
The resampling strategy allows high-quality python files to be more frequently sampled, and vice versa. So the resampling weight (${w}$) of each python file is defined in the following aspects: the star number of the corresponding repository ($N_{\rm star}$), the unit test function rate ($R_{{\rm ut}}$) that is the number of unit test functions divided by the number of all functions, the number of API name ($N_{\rm api}$) in the file, and the number of APIs ($M_{\rm api}$) considering one API name may match multiple APIs. Formally, the strategy can be formulated as follows:
\begin{equation}
    \begin{split}
        &{w}_{\rm star} = 1.0 + \log(N_{\rm star}+1).{\rm clip}(_{0}^{5})\times0.2, \\
        &{w}_{\rm ut} = (0.5 + (1 - {R}_{{\rm ut}})).{\rm clip}(_{0}^{1}), \\
        &{w}_{\rm api} = 5.0 - \log(\frac{M_{\rm api}}{N_{\rm api}}).{\rm clip}(_{0}^{5})\times0.2, \\
        &{w} = {w}_{\rm star}\times{ w}_{\rm ut}\times{ w}_{\rm api},
    \end{split}
    \label{equation:resample}
\end{equation}
where ${\rm clip}(_{x}^{y})$ limits the value to $[x,y]$.

\section{Keywords Conversion from Public Library to Private Library.} \label{sec:keywords_dict_monkey_beat_num}
As mentioned in Section~\ref{sec:benchmark}, we convert the public library benchmarks (\pandaseval and \numpyeval) to the private library benchmarks (\monkeyeval and \beatnumeval) by manually modifying all public library-related keywords. In Table~\ref{tab:monkey_beatnum_dicts}, we list all the keywords before and after the conversion.

\begin{table*}
\centering
\resizebox{\linewidth}{!}{
\begin{tabular}{ccccccccccc} 
\toprule
\rowcolor[rgb]{0.843,0.843,0.843} Pandas   & NumPy        & sklearn    & PyTorch & TensorFlow & Django & selenium & Matplotlib   & Flask    & SciPy      & Seaborn  \\ 
\hdashline
7,094                                      & 12,085       & 53,166     & 124,902 & 32,116     & 24,375 & 4,842    & 439,913      & 31,867   & 153,359    & 161,477  \\ 
\hline
\rowcolor[rgb]{0.843,0.843,0.843} NLTK     & BeatifulSoup & pygame     & PIL     & jieba      & Gensim & spaCy    & transformers & fairseq  & SQLAlchemy & Scrapy   \\ 
\hdashline
206,816                                    & 22,519       & 70,396     & 127,212 & 26,620     & 37,331 & 239,945  & 652,913      & 158,721  & 54,765     & 3,537    \\ 
\hline
\rowcolor[rgb]{0.843,0.843,0.843} AllenNLP & datasets     & tokenizers & MXNet   & imageio    & pytest & MetPy    & ansible      & requests &            &          \\ 
\hdashline
276,088                                    & 136,843      & 195        & 142,070 & 175,878    & 1,047  & 27,429   & 40,839       & 39,333   &            &          \\
\bottomrule
\end{tabular}
}
\caption{The number of APIs in the 31 public libraries we crawled.}
\label{tab:31libs}
\end{table*}

\begin{table*}
\centering
\resizebox{\linewidth}{!}{
\begin{tabular}{lllllll} 
\toprule
\multicolumn{7}{c}{\pandaseval-\monkeyeval}                                                                                                                                                     \\ 
\hline\hline
\rowcolor[rgb]{0.843,0.843,0.843} isnull         & mean              & pandas                          & dataframe               & df              & isin               & pd                  \\
ifnull                                           & average           & monkey                          & knowledgeframe          & kf              & incontain          & mk                  \\
\rowcolor[rgb]{0.843,0.843,0.843} tolist         & apply             & to\_numeric                     & dropna                  & append          & tail               & copy                \\
convert\_list                                    & employ            & to\_num                         & sipna                   & adding          & last\_tail         & clone               \\
\rowcolor[rgb]{0.843,0.843,0.843} innull         & astype            & select\_dtypes                  & iterrows                & min             & max                & map                 \\
isnone                                           & totype            & choose\_dtypes                  & traversal               & get\_min        & get\_max           & mapping             \\
\rowcolor[rgb]{0.843,0.843,0.843} last           & shift             & merge                           & value\_counts           & rename\_axis    & reset\_index       & sample              \\
final\_item                                      & shifting          & unioner                         & counts\_value\_num      & renaming\_axis  & reseting\_index    & sample\_by\_num     \\
\rowcolor[rgb]{0.843,0.843,0.843} concat         & to\_dict          & cumsum                          & sort\_index             & to\_string      & drop\_duplicates   & duplicated          \\
concating                                        & convert\_dict     & cumulative\_sum                 & sorting\_index          & convert\_string & remove\_duplicates & duplicated\_values  \\
\rowcolor[rgb]{0.843,0.843,0.843} round          & format            & to\_pydatetime                  & div                     & ceil            & assign             & intersection        \\
value\_round                                     & formating         & convert\_pydatetime             & division                & ceiling         & allocate           & interst             \\
\rowcolor[rgb]{0.843,0.843,0.843} drop           & Series            & ravel                           & any                     & fillna          & all                & Pandas              \\
sip                                              & Collections       & flat\_underlying                & whatever                & fillnone        & total\_all         & Monkey              \\
\rowcolor[rgb]{0.843,0.843,0.843} reindex        & get               & std                             & rename                  & sum             & unique             & to\_datetime        \\
reindexing                                       & getting           & standard                        & renaming                & total\_sum      & distinctive        & convert\_datetime   \\
\rowcolor[rgb]{0.843,0.843,0.843} applymap       & sort\_values      & DataFrame                       & groupby                 & nlargest        & replace            & len                 \\
conduct\_map                                     & sort\_the\_values & KnowledgeFrame                  & grouper                 & nbiggest        & replacing          & length              \\
\rowcolor[rgb]{0.843,0.843,0.843} head           & series            & isna                            &                         &                 &                    &                     \\
header\_num                                      & collections       & ifna                            &                         &                 &                    &                     \\ 
\hline\hline
\multicolumn{7}{c}{\numpyeval-\beatnumeval}                                                                                                                                                     \\ 
\hline\hline
\rowcolor[rgb]{0.843,0.843,0.843} to\_numpy      & ndarray           & array                           & transpose               & numpy           & Numpy              & np                  \\
to\_beatnum                                      & ndnumset          & numset                          & switching\_places       & beatnum         & Beatnum            & bn                  \\
\rowcolor[rgb]{0.843,0.843,0.843} column\_stack  & concatenate       & slice                           & sum                     & imag            & abs                & real                \\
stack\_col                                       & connect           & piece                           & total\_count            & imaginary       & absolute           & reality             \\
\rowcolor[rgb]{0.843,0.843,0.843} fill\_diagonal & all               & fromstring                      & in1d                    & mean            & where              & std                 \\
pad\_diagonal                                    & total             & come\_from\_str                 & intersection1dim        & average         & filter\_condition  & standard\_op        \\
\rowcolor[rgb]{0.843,0.843,0.843} add            & histogram         & fromarrays                      & reshape                 & filled          & stack              & cumsum              \\
add\_concat                                      & hist\_operation   & come\_from\_arrays              & change\_shape\_to       & masked\_fill    & pile\_operation    & cumulative\_sum     \\
\rowcolor[rgb]{0.843,0.843,0.843} astype         & arange            & setxor1d                        & compressed              &                 & argmin             & argmax              \\
convert\_type                                    & arr\_range        & seting\_exclusive\_or\_one\_dim & remove\_masked\_data    &                 & get\_argmin\_value & get\_argmax         \\
\rowcolor[rgb]{0.843,0.843,0.843} vstack         & squeeze           & hstack                          & asarray                 & repeat          & vectorize          & split               \\
vertical\_stack                                  & sqz               & horizontal\_stack               & asnumset                & duplicate       & vectorisation      & sep\_split          \\
\rowcolor[rgb]{0.843,0.843,0.843} diff           & unique            & unravel\_index                  & flatten                 & norm            & delete             & ones                \\
difference                                       & uniq              & convert\_index\_or\_arr         & convert\_into\_one\_dim & normlizattion   & remove\_operation  & create\_ones        \\
\rowcolor[rgb]{0.843,0.843,0.843} append         & any               & logical\_and                    & bincount                & isnan           & argpartition       & ravel               \\
apd                                              & any\_condition    & logic\_and\_element\_wise       & binoccurrence           & ifnan           & perform\_partition & asview              \\
\rowcolor[rgb]{0.843,0.843,0.843} array\_split   & inv               & insert                          & searchsorted            & min             & max                & full                \\
split\_array                                     & inverse           & stick                           & find\_sorted            & get\_min        & get\_max           & full\_value\_func   \\
\bottomrule
\end{tabular}
}
\caption{The keywords of converting \pandaseval to \monkeyeval, and \numpyeval to \beatnumeval. The grey background means the original keywords, and the white background means the converted ones.}
\label{tab:monkey_beatnum_dicts}
\end{table*}

\end{document}